\documentclass[runningheads]{svmult}

\usepackage{makeidx}   
\usepackage{graphicx}  
\usepackage{subeqnar}  
\usepackage{multicol}  
\usepackage{physprbb}  
\makeindex             

\begin{document}
\title*{Mass/Light Variations with Environment}
\toctitle{Mass/Light Variations with Environment}
%
%
\titlerunning{Mass/Light Variations}
%
\author{R. Brent Tully\inst{1}}
\authorrunning{R. Brent Tully}
%
%
\institute{University of Hawaii, Honolulu HI 96822 USA}

\maketitle              

\begin{abstract}
In the local part of the universe, a majority of the total mass is in only a
small number of groups with short dynamical times.  The Virgo Cluster dominates
far out of proportion to its contribution in light.  Ninety percent of the
total mass is in groups with $12.5 < {\rm log} M/M_{\odot} < 15$.
\end{abstract}

\section{Introduction}
It would hardly be surprising if the relationship between mass and light varies
with environment, even if structures form from an invariant baryon to dark
matter fractionation.  Dark matter could be
more dispersed than the baryons that are manifested in stars or detectable
gas.  This possible differentiation would create a `bias' between what is seen
and what exists [1].  Recent modelling [2,3] suggests there could be a complex
relationship betwen mass and light, 
that dark matter may be underrepresented by light at both extremes of low 
density and high density.
Mass-to-light may grow with scale around galaxies
to an asymptotic limit [4].
This paper presents observational evidence from the motions of galaxies that
there are mass-to-light differences with environment.

\section{Galaxy Groups}

Dark halos extend beyond the observed baryonic components of galaxies so
there cannot be a sufficient understanding of the distribution of dark matter
from studies of the internal kinematics of galaxies.  The motions of galaxies 
relative to 
each other provide a probe of the distribution of matter on the scales of the
separations between objects.

This discussion will begin with a new look at old data.  Masses can be 
estimated for groups of galaxies using the virial theorem.  To be 
strictly applicable, the group should be relaxed.  However, within a possible 
systematic of a factor of two, the virial theorem gives an estimate of group
masses as long as the basic assumption is met that the group is bound.

Candidate groups can be found among the structures defined by a dendogram
analysis [5]. 
Within a sample of $N$ galaxies, the pair is selected with the extreme of an 
appropriate property, say,
the largest product luminosity / separation$^2$.  This pair is then
considered as a single unit and the procedure is repeated.  After $N-1$ steps,
all galaxies in the sample are merged.
At each merger step, a luminosity density can be characterized by the sum of
the light of the components divided by the cube of the separation at this step.
Now the timescale for separation of an overdensity from the cosmic expansion
and its collapse depends on the inverse square root of the mass overdensity.  
If there is a constant relationship between mass and light, then luminosity
is a standin for mass.  Then, in a plot of the
merger dendogram as a function of luminosity density there would be a 
discrete cut that would distinguish entities that could have collapsed within
the age of the universe.

A constant relationship between mass and light remains to be demonstrated and
it would not be appropriate to specify such a relationship 
since this is to be a product of the study.  However an independent observable
monitors the probability that a candidate group is bound: the crossing time - a
characteristic dimension for the group divided by the velocity dispersion.
Assume a cut on the luminosity density dendogram.  If that cut is at a low
density, there would be a lot of entities above the cut with crossing times 
longer than the age of the universe.  If the cut is at a very high density,
the entities that survive the cut would all have crossing times much shorter
than the age of the universe.  Like Goldilocks, we try to find a cut that is
just right, characterized by crossing times that scatter up to, but not above,
the age of the universe.  The details of the choice of cut and consequent
properties of a sample of groups within the Local Supercluster have been 
discussed [5].
The Nearby Galaxies Catalog [6] includes these group
affiliations.  This catalog also records entities as `associations' if they
satisfy a luminosity density threshold an order of magnitude lower than the 
cut that specifies groups.  This more lax prescription saves a more extended
list of group candidates, a matter of interest later in the discussion.


\begin{figure}[]
\begin{center}
\includegraphics[width=.9\textwidth]{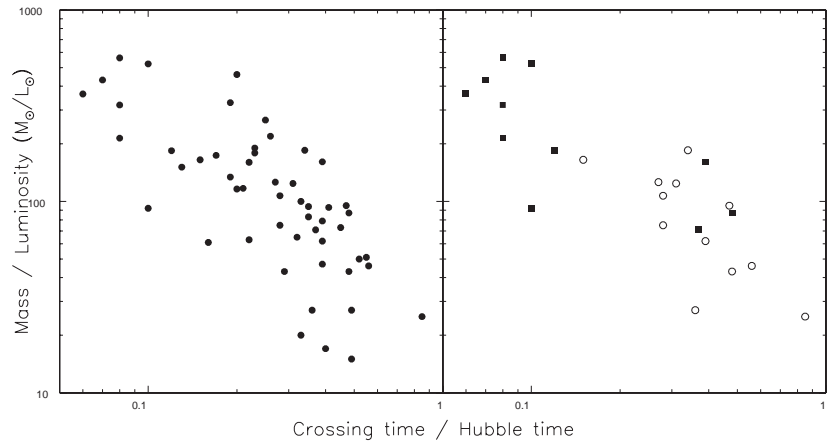}
\end{center}
\caption[]{Correlation between dynamical time and $M/L_B$. {\it Left panel:}
groups with at least 5 identified members. {\it Right panel:} solid squares
identify groups with a majority of types
E-S0-Sa; open circles identify groups with a majority of types Sab-Irr}
\label{fig1}
\end{figure}

A fresh look at this old material reveals that there are strong correlations
between mass-to-blue-light ratio ($M/L_B$) and either crossing time or 
morphology.
These correlations are seen in Figure~1 where $M/L_B$ is plotted against
crossing time for a 
sample of 49 groups with at least 5 members that lie within $25 h_{75}^{-1}$
Mpc.
In the panel at the right, only
groups with at least 6 known members with $M_B<-17$ are considered
(here, H$_{\circ}=75$ km s$^{-1}$ Mpc$^{-1}$ and $h_{75}={\rm H}_{\circ}/75$).

Fig.~1 reveals that groups predominantly composed of early type galaxies
tend strongly to have short 
crossing times and high $M/L_B$.  Groups of predominantly late type galaxies
inevitably have longer
crossing times and lower $M/L_B$.  The range in $M/L_B$ values extends over 
more than a decade.  The trend in Fig.~1 has too big an amplitude be explained
by correlated
errors in velocity dispersion in the calculation of $M/L_B$ and crossing time.
The correlation between group morphology and crossing time rests on
independent information.  
The same information is seen in
Figure~2.  The symbols again denote group morphology.  There is a trend with 
group mass, with a larger fraction of early type groups toward higher masses.
The consequence is a roll off toward a shallow slope between mass
and light at high mass.  It is seen, though, that at intermediate masses around $10^{13}~M_{\odot}$ there is a substantial range in $M/L_B$ values.  The
variance is strongly correlated with both crossing time and group morphology.

\begin{figure}[]
\begin{center}
\includegraphics[width=1.\textwidth]{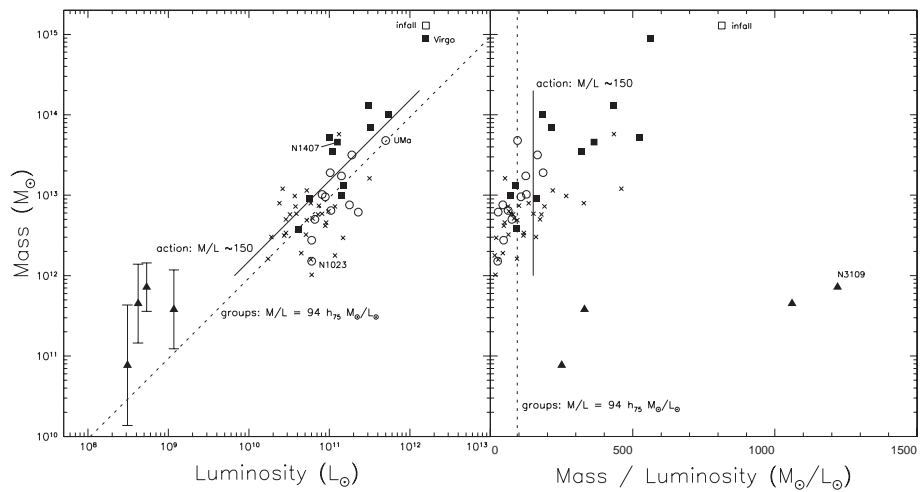}
\end{center}
\caption[]{In the left panel, group mass is plotted against group luminosity.
In the right panel, the same mass is plotted against mass-to-light ratio.
Filled squares and open circles represent the same groups seen in the right
panel of Fig 1 and the crosses represent the smaller groups that fill out the 
left
panel of Fig 1.  The 4 triangles with error bars locate the
groups of 
dwarfs discussed in section 4.  The dashed lines indicate the mean $M/L$ value
found for groups [5], the solid lines indicate the mean field result
from numerical action models [9], and the open box is based 
on the mass requirement for the Virgo Cluster that follows from the numerical
action models
 }
\label{fig2}
\end{figure}

\section{Numerical Action Modelling}

The distribution of dark matter on scales larger than groups is recorded in 
the slosh of galaxies about Hubble flow on scales of megaparsecs to tens of
megaparsecs.  The general properties of the gravitational field can be 
recovered through the construction of plausible galaxy orbits with mixed
boundary condition constraints [7,8].
It is to be appreciated that, while to date the specifics of individual orbits
are poorly constrained by this modelling, mean mass densities are tightly
constrained.  In the general field,
$M/L_B\sim 150~M_{\odot}/L_{\odot}$ is found from this modelling [9]
a result indicated by the solid lines in Fig.~2.  This $M/L_B$ value is 
50\% larger than the mean value derived from the group analysis which may be
reasonable given the reference to scales larger than the domain of bound 
groups.

The action modelling of the Local Supercluster region raised an important
issue regarding $M/L_B$ variations with environment.  There is specific 
information about the infall of galaxies toward the Virgo Cluster.  Galaxies
are entering the cluster with 1-D velocities in excess of 1500~km~s$^{-1}$
and the current zero-velocity shell with respect to the cluster is at about
$28^{\circ}$ radius from the cluster center.  It is impossible to provide an
adequate description of this infall with a mass assigned to the Virgo Cluster
in accordance with the low $M/L_B$ value found overall.  The infall pattern
requires a mass of
$1.4\times 10^{15} h_{75}^{-1}~M_{\odot}$ and 
$M/L_B\sim 900 h_{75}^{-1}~M_{\odot}/L_{\odot}$ 
[9].  These values are large but the mass
relates to the cluster on a scale in excess of 2 Mpc radius.  This datum is
recorded by the open square in Fig.~2.  
The key constraints are (a) a high mass is required in the cluster to explain the
high infall velocities in the immediate vicinity of the cluster, but (b)
if there were a significant amount of additional mass betwen us and the
cluster then we would have a higher retardation from Hubble flow.

\section{Groups of Dwarfs}

It was mentioned that the group analysis [5] also identified entities called
associations that pass a luminosity density cut an order of magnitude fainter
than the group cut.  There are a small fraction of entities identified this
way that deserve special attention.  There are a few associations comprised
of only dwarf galaxies which would easily have been called groups if only
separations were considered (ie, luminosity not a factor).  Velocity 
dispersions for these associations are extremely low, of order 20~km~s$^{-1}$, 
so if these entities
are bound then masses are modest, of order $10^{11}~M_{\odot}$.  However there
is little light so $M/L_B$ values range from 
$\sim 250$ to $1200~M_{\odot}/L_{\odot}$.

These proposed groups have been discussed [10] in the context of
a plausible galaxy formation scenario.  The details of the formation mechanism
will not be pursued here but the generic possibility is raised that there
could be many low mass dark halos with little or no accompanying stars or gas.
It would follow from this hypothesis that there could be environments with 
{\it some} star formation, enough to provide probes of the potential, but
still so little that $M/L_B$ is large.

\begin{figure}[]
\begin{center}
\includegraphics[width=.6\textwidth]{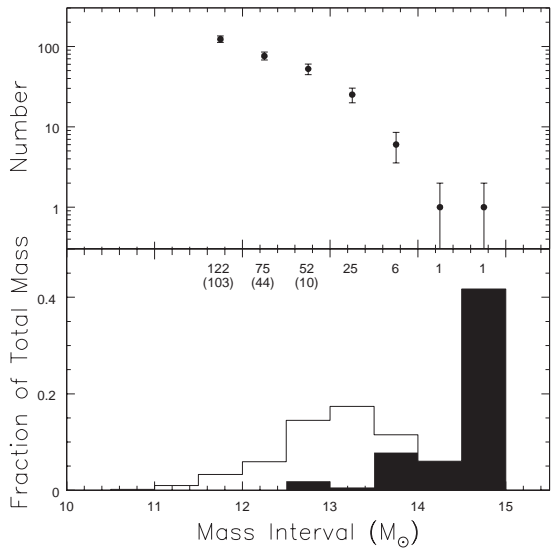}
\end{center}
\caption[]{{\it Lower panel:} fraction in half decade mass intervals of the 
total mass associated 
with local groups and galaxies.  Filled histogram:
14 groups with crossing times less than 0.2H$_0^{-1}$. Open histogram: 
all other groups and individual galaxies.
Numbers of groups plus individual galaxies per bin are
given across the top of the panel (numbers of individual galaxies in brackets).
The sample is reasonably complete above ${\rm log}M = 11.5$. {\it Upper panel:}
number of groups plus individual galaxies per mass interval}
\label{fig3}
\end{figure}

The four candidate dwarf groups that are 
represented at the low mass end of Fig~2 are identified and details on their 
properties are given in reference [10].  The critical issue is whether the 
proposed groups are 
bound because, if so, then $M/L_B$ values are surely high.  It is to be noted
that these 4 candidate groups are all within 5~Mpc and within this distance
beyond the Local Group (restricting to high latitudes) there are only 4
`normal' groups with luminous galaxies.  There is the prospect that bound 
groups of dwarfs are numerically common, though they may contribute only a
small fraction of the total mass of the universe.  One could suspect that
dark matter structures devoid of stars would also exist, which would be in
line with the original `biasing' idea.  However the action analysis suggests
that such structures would not contribute too much to the overall mass
inventory.

\section{Conclusions}

Figure~3 shows the
inventory of the clustered mass in the local region.
Mass contributions are summed over all groups and individual 
galaxies within a distance of $25 h_{75}^{-1}$~Mpc and with
$\vert b \vert > 30^{\circ}$ (distances based on a numerical action kinematic
model).  In the case of groups, masses come from application of the
virial theorem.  In the instances of pairs or triples where the observed 
dispersion
in velocities is dominated by measurement uncertainties, or in the case of
single galaxies, masses are inferred assuming 
$M/L_B = 100 M_{\odot}/L_{\odot}$,
a round off of the mean result for groups represented by the dotted lines in
Fig 2.  The volume contains a total
luminosity of $1.0 \times 10^{13} L_{\odot}$ in cataloged galaxies and a
total mass of $2.1 \times 10^{15} M_{\odot}$ (only 10\% of this mass total is
inferred from the $M/L$ assumption).  Within this nearby volume, there 
should be reasonable completion down to
$0.1 L_B^{\star}$.  The overall $M/L \sim 200 M_{\odot}/L_{\odot}$ is 
consistent with $\Omega_m \sim 0.2$ [9].  The mean density
in this local region is $4.4 \times 10^{-30}h_{75}^2$ gm/cm$^3$, indicating
this region has twice the cosmic mean density if $\Omega_m = 0.2$.  Hence our 
census accounts for all the
mass anticipated by the numerical action dynamical modelling.
There are two striking points to note with Fig 3.  One point is that 90\% of 
the mass is in bound entities with ${\rm log}M > 12.5$.  The other point is that
the Virgo Cluster, with 15\% of the light in this volume, has over 40\% of the 
mass!  Of course, the statistics
above ${\rm log}M = 14$ are inadequate with this local sample (Fornax Cluster is
the second massive entity).

The information provided by groups about the distribution of dark matter
on scales of hundreds of kiloparsecs and the information provided by galaxy 
flows about dark matter on scales up to tens of megaparsecs
give rise to a consistent picture. $M/L_B$ values in dynamically evolved
regions can be an order of magnitude higher than in the great majority of
places that are dynamically young.  The evidence from both galaxy 
motions and an inventory of groups 
suggests $M/L_B\sim 200~M_{\odot}/L_{\odot}$ overall, consistent with 
$\Omega_{matter}\sim 0.2$.  Similar results are found from wide field 
weak lensing studies [11].
Within the high latitude, inner 25 Mpc region of the group sample, roughly 
60\% of 
the mass is in 14 low crossing time groups dominated by early galaxy types
that contribute a quarter
of the light.  We seem to live in a curious universe 
where a majority of the clumped matter is in the modest percentage of 
locations with crossing times $<2$ Gyr.  

Groups containing familiar luminous galaxies all have masses above 
$\sim 10^{11}~M_{\odot}$.  Candidate groups of only dwarf galaxies are 
identified with masses near this $10^{11}~M_{\odot}$ limit. These associations
are probably bound, whence they would contain mostly dark matter.  It is
reasonable to speculate that there are low mass halos without any stars or 
gas.  However most of the mass of the universe seems to be in collapsed 
regions with $10^{12.5}-10^{15}~M_{\odot}$.

%

\end{document}